\documentclass[11pt]{article}
\usepackage{geometry}                % See geometry.pdf to learn the layout options. There are lots.
\usepackage{graphicx}
\usepackage{amssymb}
\usepackage{amsmath}
\usepackage{amsfonts}
\usepackage{graphicx}
\usepackage{epstopdf}
\usepackage{color}
\usepackage{cite}
\usepackage{ulem}
\usepackage{soul}

\newcommand{\be}{\begin{equation}}
\newcommand{\ee}{\end{equation}}
\newcommand{\bea}{\begin{eqnarray}}
\newcommand{\eea}{\end{eqnarray}}
\newcommand{\beas}{\begin{eqnarray*}}
\newcommand{\eeas}{\end{eqnarray*}}
\newcommand{\avg}[1]{\left\langle{#1}\right\rangle}

\newcommand{\w}{\vec{w}}

\usepackage{float}
\usepackage[titletoc,title]{appendix}

\usepackage{parskip}

\title{Bias-variance trade-off in portfolio optimization under Expected Shortfall with $\ell_2$ regularization}
\date{}
\author{G\'abor Papp$^1$, Fabio Caccioli$^{2,3}$ and Imre Kondor$^{4,5,6}$ \\
{\it 1 - E\"otv\"os Lor\'and University, Institute for Physics, Budapest, Hungary } \\
{\it 2 - University College London, Department of Computer Science,} \\
{\it London, WC1E 6BT, UK} \\
{\it 3 - Systemic Risk Centre, London School of Economics and Political Sciences, London, UK}\\
{\it 4 - Parmenides Foundation, Pullach, Germany}\\
{\it 5 - London MathematicalLaboratory, London, UK} \\
{\it 6 - Complexity Science Hub, Vienna, Austria} \\
}

\begin{document}
\bibliographystyle{unsrt}

%\begin{abstract} 
%\end{abstract} 
\maketitle

\begin{abstract} 
The optimization of a large random portfolio under the Expected Shortfall risk measure with an
$\ell_2$ regularizer is carried out by analytical calculation. The regularizer reins in the large
sample fluctuations and the concomitant divergent estimation error, and eliminates the phase
transition where this error would otherwise blow up.  In the data-dominated region, where the number $N$
of different assets in the portfolio is much less than the length $T$ of the available time series, the
regularizer plays a negligible role even if its strength $\eta$ is large, while in the
opposite limit, where the size of samples is comparable to, or even smaller 
than the number of assets, the optimum is almost entirely determined by the regularizer. We construct the contour map 
of estimation error on the $N/T$ vs. $\eta$ plane and find that for a given value of the estimation error the gain in $N/T$ 
due to the regularizer can reach a factor of about 4 for a sufficiently strong regularizer. 

\end{abstract}

\section{Introduction}
The current international market risk regulation measures risk in terms of Expected Shortfall (ES) \cite{basle2016Minimum}.
In order to decrease their capital requirements, financial institutions have to optimize the ES of their trading book. 

Optimizing a large portfolio may be difficult under any risk measure, but becomes
particularly hard in the case of Value at Risk (VaR) and Expected Shortfall (ES) 
that discard a large part of the data except those at or above a high quantile. Without some kind of
regularization, this leads to a phase transition at a critical value $r_c$ of the ratio $r=N/T$ where $N$ is the dimension 
of the portfolio (the number of different assets or risk factors) and $T$ is the sample size (the length of the available time series). 
This critical ratio depends on the confidence level $\alpha$ that determines the VaR threshold above which the losses are 
to be averaged to obtain the ES. Beyond $r_c$  it is impossible to carry out the optimization, and upon approaching 
this critical value from below the estimation error increases without bound. 

The estimation error problem of portfolio selection has been the subject of a large number of works, \cite{jorion1986Bayes,frost1986AnEmpirical,ledoit2003Improved,ledoit2004AWell,ledoit2004Honey,golosnoy2007Multivariate,brodie2009Sparse,laloux1999Noise,laloux2000Random,plerou1999Universal,plerou2000ARandom} are but a small selection from this vast literature. 
The critical behavior and the locus of the phase boundary separating the region where the optimization is feasible from the one where it is not has also been studied in a series of papers \cite{ciliberti2007On,ciliberti2007Risk,varga2008TheInstability,caccioli2013Optimal,caccioli2016liquidity}. 

In the present note we discuss the effect of adding an $\ell_2$ regularizer to the ES risk
measure. As noted in \cite{takeda2008nu} and \cite{still2010Regularizing}, the optimization problem
so obtained is equivalent to one of the variants of support vector regression ($\nu$-SVR) \cite{perez2003Extension}, therefore its study
is of interest also for machine learning, independently of the portfolio optimization
context. Concerning its specific financial application, $\ell_2$ regularization may have two
different interpretations. First, $\ell_2$ has the tendency to pull the solution towards the
diagonal, where all the weights are the same. As such, $\ell_2$ represents a diversification pressure
\cite{bouchaud2003Theory,gabor1999Portfolios} that may be useful in a situation where, e.g., the
market is dominated by a small number of blue chips. Alternatively, the portfolio manager may wish
to take into account the market impact of the future liquidation of the portfolio already at its
construction. As shown in \cite{caccioli2016liquidity}, market impact considerations lead one to
regularized portfolio optimization, with $\ell_2$ corresponding to linear impact.

In this paper we carry out the optimization of $\ell_2$-regularized ES analytically in the special
case of i.i.d.\;Gaussian distributed returns, in the limit where both the dimension and the sample
size are large, but their ratio $r=N/T$ is fixed. The calculation will be performed by the method of
replicas borrowed from the statistical physics of disordered systems \cite{mezard1987Spin}. The
present work extends a previous paper \cite{caccioli2015Portfolio} by incorporating the
regularizer. By preventing the phase transition from taking place, the regularizer
fundamentally alters the overall picture (in this respect, the role of the regularizer is analogous
to that of an external field coupled to the order parameter in a phase transition). 
As the technical details of the replica calculation have been laid out in \cite{ciliberti2007On} and, in a somewhat
different form, in \cite{caccioli2016liquidity}, we do not repeat them here. Instead we just recall the setup of
the problem and focus on the most important result: the relative estimation error (closely related to the out-of-sample estimator of ES) as a function 
of $r=N/T$ and the strength of the regularizer $\eta$.

Our results exhibit a clear distinction between the region in the space of
parameters where data dominate and the regularizer plays a minor role, from the one where the data
are insufficient and the regularizer stabilizes the estimate at the price of essentially suppressing
the data. Thereby, our results provide a clean and explicit example of what statisticians call the bias-variance trade-off  
that lies at the heart of the regularization procedure. We find that the transition between
the data-dominated regime and the bias-dominated one is rather sharp, and it is only around this transition that an actual
trade-off takes place. Following the curves of fixed estimation error on the $r -\eta$ plane we can see that $r$ increases with 
increasing $\eta$ by a factor of approximately 4. Beyond this point the contour lines turn back and we go over onto a branch of the contour line where the 
optimization is determined by the regularizer rather than the data. 
The plan of the rest of the paper is as follows. In Sec. 2 we present the formalism, in Sec 3 display our results and in Sec. 4 draw our conclusions. 

\section{The optimization of regularized ES}

The simple portfolios we consider here are linear combinations of $N$ risk factors, 
with returns ${x_i}$ and weights ${w_i},~i=1,2,...,N$:

\be
X = \sum_{i=1}^N w_i x_i
\ee

The weights will be normalized such that their sum is $N$, instead of the customary 1:

\be\label{eqBudgetConstraint}
\sum_{i=1}^N w_i = N.
\ee

The motivation for choosing this normalization is that we wish to have weights of order unity, rather
than $1/N$, in the limit $N\to\infty$. Apart from the budget constraint, the weights will not be subject 
to any other condition. In particular, they can take any real value, that is we are allowing unlimited short
positions. We do not impose the usual constraint on the expected return on the portfolio either, so 
we are looking for the global minimum risk portfolio. This setup is motivated by simplicity, but we note 
that tracking a benchmark requires exactly this kind of optimization to be performed.

The probability for the loss $\ell(\{w_i\},\{x_i\})=-X$  to be smaller than a threshold $\ell_0$ is: 

\begin{equation*}
P(\{w_i\},\ell_0) = \int \prod_{i=1}^N d x_i p(\{x_i\}) \theta\left(\ell_0- \ell(\{w_i\},\{x_i\})\right)
\end{equation*}

where $p(\{x_i\})$ is the probability density of the returns, and $\theta(x)$ is the Heaviside function: $\theta(x) = 1$ for $x> 0$, and zero otherwise. 
The Value at Risk (VaR) at confidence level $\alpha$ is then defined as:

\be
{\rm VaR}_\alpha(\{w_i\}) = {\rm min} \{\ell_0 : P(\{w_i\},\ell_0) \ge\alpha\}.
\ee

Expected Shortfall is the average loss beyond the VaR quantile:

\be\label{eqESDefinition}
{\rm ES}(\{w_i\}) = \frac{1}{1-\alpha}\int \Pi_i dx_i p(\{x_i\}) \ell(\{w_i\},\{x_i\}) \theta(\ell(\{w_i\},\{x_i\}) - {\rm VaR}_\alpha(\{w_i\}) ).
\ee

Portfolio optimization seeks to find the optimal weights that make the above ES minimal subject to the budget 
constraint \eqref{eqBudgetConstraint}. Instead of dealing directly with ES, Rockafellar and Uryasev 
\cite{rockafellar2000Optimization} proposed to minimize the related function

\be
F_\alpha ( \{w_i\},\epsilon) = \epsilon +\frac{1}{1-\alpha}\int \Pi_i dx_i p(\{x_i\})\left[\ell(\{w_i\},\{x_i\})-\epsilon\right]^+
\ee

over the variable $\epsilon$ and the weights $w_i$:

\be
{\rm ES}(\{w_i\}) = {\rm min}_{\epsilon} F_\alpha ({\{w_i\},\epsilon}),
\ee

where $[x]^+ =(x+|x|)/2$.

The probability distribution of the returns is not known, so one can only sample it,
and replace the integral in (4) by time-averaging over the discrete observations. Rockafellar and
Uryasev \cite{rockafellar2000Optimization} showed that the optimization of the resulting loss
function can be reduced to the following linear programming task: Minimize the cost function

\be\label{eqCostFunction}
E(\epsilon,\{u_t\} )= (1-\alpha) T \epsilon + \sum_{t=1}^T u_t
\ee
under the constraints
\begin{equation*}
u_t  \ge 0~~\forall~t,
\end{equation*}
\be
u_t +  \epsilon+\sum_{i=1}^N x_{it} w_i \ge 0~~\forall~t,
\ee
\begin{equation*}
{\rm and}~~~\sum_i w_i =N.
\end{equation*}

It is convenient to introduce the regularizer at this stage, by adding the $\ell_2$-norm to the loss function \cite{caccioli2016liquidity}:

\begin{eqnarray}
&&\min_{\w, {\vec u}, \epsilon}  \left[ (1-\alpha) T\epsilon+\sum_{t=1}^T u_{t}+{\eta } \sum_iw_i^2\right],
\label{newPO} \\
&{\rm s.t.}\;\;\; & \vec{w}\cdot\vec{x}_t + \epsilon + u_t \geq 0; \;\;\; u_t \geq 0; \;\;\; \forall t, \label{constraints} \\
&& \sum_i w_i = N,
\end{eqnarray}
where $\epsilon$ and $\vec u$ are auxiliary variables, and the coefficient $\eta$ sets the strength of the regularization. 

As the constraint on the expected return has been omitted, we are seeking the global optimum of the portfolio here.
If the returns $x_{it}$ are i.i.d.\;Gaussian variables and $N,T \to\infty$ with $r=N/T$ fixed, the method of replicas 
allows one to reduce the above optimization task in $N+T+1$ variables to the optimization of a cost function 
depending on only six variables, the so-called order parameters \cite{ciliberti2007On,caccioli2016liquidity}:

\begin{eqnarray}
\label{free_energy}
F({\lambda},{\epsilon},{q}_0,\Delta, {\hat{q}}_0,\hat{\Delta})&=&
{\lambda} +\frac{1}{r} (1-\alpha)\epsilon -\Delta{\hat{q}}_0-\hat{\Delta}{q}_0\\
\nonumber &+& \langle {\rm min}_w \left[V(w,z)\right]\rangle_z
 +\frac{\Delta}{2r\sqrt{\pi}}\int_{-\infty}^{\infty}ds\: e^{-s^2}
g\biggl({\frac{\epsilon}{\Delta}}+s \sqrt{\frac{2{q}_0}{\Delta^2}} \biggr),
\end{eqnarray}

where  
\begin{equation}
\label{potential}
V(w,z)=\hat{\Delta} w^2 -{\lambda} w -z w\sqrt{-2{\hat{q}}_0} + \eta w^2,
\end{equation}
$\langle \cdot \rangle_z$ is an average over the standard normal variable $z$,
and
\begin{equation}
    g(x)= \left\{ \begin{array}{cc} 0 ,&    x\ge 0\\
    x^2 , &  -1\le x\le 0\\
    -2 x-1, & x<-1 %\risi{\: .}
     \end{array} \right. %\rdel{.}
\end{equation}

One can readily see that the stationarity conditions are:

\be
1=\avg{w^*}_{z}\label{spBudget}
\ee
\be
\label{EqFirstOrder2}
(1-\alpha)+\frac{1}{2\sqrt{\pi}}\int_{-\infty}^\infty ds\: e^{-s^2}
g'\biggl(\frac{\epsilon}{\Delta}+s\sqrt{\frac{2{q}_0}{\Delta^2}}\biggr)=0
\ee
\be
\hat{\Delta} - \frac{1}{2r\sqrt{2\pi{q}_0}}\int_{-\infty}^\infty ds\: e^{-s^2}s 
g'\biggl(\frac{\epsilon}{\Delta}+s\sqrt{\frac{2{q}_0}{\Delta^2}}\biggr)=0
\ee

\be
-{\hat{q}}_0-2\frac{\hat{\Delta}{q}_0}{\Delta} + \frac{1}{2r\sqrt{\pi}}\int_{-\infty}^\infty ds\: 
e^{-s^2}g\biggl(\frac{\epsilon}{\Delta}+s\sqrt{\frac{2{q}_0}{\Delta^2}}\biggr)+\frac{(1-\alpha)}{r}\frac{{\epsilon}}{\Delta}=0
\ee

\be
\Delta=\frac{1}{\sqrt{-2{\hat{q}}_0}}\avg{w^* z}_{z}\label{spDelta}
\ee

\be
{q}_0=\avg{{w^*}^2}_{z}\label{spQ},
\ee
where the variable $w^*$ is that value of the weight that minimizes the ``potential'' $V$ in \eqref{potential}.
(The prime means derivative with respect to the argument.) 

Three of the order parameters are easily eliminated and the integrals can be reduced to the error function and its integrals by repeated integration by parts, as in \cite{caccioli2015Portfolio}. Finally, one ends up with three equations to be solved:

\be
 \label{equationPhizd}
  r \left( 1  - 2\eta \Delta \right) = \Phi\left( \frac{\Delta+\epsilon}{\sqrt{q_0}} \right) - \Phi\left(\frac{\Delta+\epsilon}{\sqrt{q_0}}\right)
\ee

\be
 \label{equationPsizd}
  \alpha =  \frac{\sqrt{q_0}}{\Delta} \left\{ \Psi\left(  \frac{\Delta+\epsilon}{\sqrt{q_0}}  \right) - \Psi\left( \frac{\Delta+\epsilon}{\sqrt{q_0}}\right)  \right\}
\ee

\be
 \label{equationWzd}
 \frac1{2\Delta^2} + \frac{\alpha}{r}\frac{\epsilon}{\Delta} + \frac{q_0}{2\Delta^2} + \frac{1}{2 r} - \frac{2\eta q_0}{\Delta^2} = \frac{q_0}{r\Delta^2}  \left\{ W\left(\frac{\Delta+\epsilon}{\sqrt{q_0}} \right) - W\left(  \frac{\Delta+\epsilon}{\sqrt{q_0}} \right) \right\} \, ,
\ee
where

\be
\Phi(x) = \frac{1}{\sqrt{2\pi}}\int_{-\infty}^x dt e^{-t^2/2},
\ee

\be
\Psi(x) = x\,\Phi(x) + \frac{1}{\sqrt{2\pi}}\:e^{-x^2/2},
\ee

\be
W(x) = \frac{x^2+1}{2}\:\Phi(x)+\frac{x}{2}\frac{1}{\sqrt{2\pi}}\:e^{-x^2/2}.
\ee

These functions are closely related to each other: $\Phi(x)$ is the derivative of $\Psi(x)$ and $\Psi(x)$ is the derivative of $W(x)$. 

As explained in \cite{caccioli2015Portfolio}, each of the three remaining order parameters in the
above set of equations, $q_0$, $\Delta$, and $\epsilon$ has a direct financial meaning: 
$\Delta$ is related to the in-sample estimator of ES (and also to the second derivative of the cost function $F$ with respect to 
the Lagrange multiplyer $\lambda$ associated with the budget constraint) and $\epsilon$ is the in-sample VaR of the 
portfolio optimized under the ES risk measure. Our present concern is the order parameter $q_0$, which is
a measure of the out-of-sample estimator of ES. As shown in \cite{caccioli2015Portfolio}, if
 ${\rm ES}_{\rm out}$ is the out-of-sample estimate of ES based on samples of size $T$, and ${\rm
ES}^{(0)}$ is its true value (that would obtain for $N$ finite and ${T\to\infty}$), then

\be
  \frac{{\rm ES}_{out}}{{\rm ES}^{(0)}} = \sqrt{q_0},
\ee

that is $\sqrt{q_0}-1$ is the relative estimation error of the out-of-sample estimate. 

The task now is to solve the stationary conditions and get the cost function by
substituting the solutions back into eq.\:\eqref{free_energy}. 
The in-sample value of Expected Shortfall is simply related to the cost function as:

\be
ES=\frac{r F}{1-\alpha}.
\ee

\begin{figure}
	\centerline{\includegraphics[width=7cm]{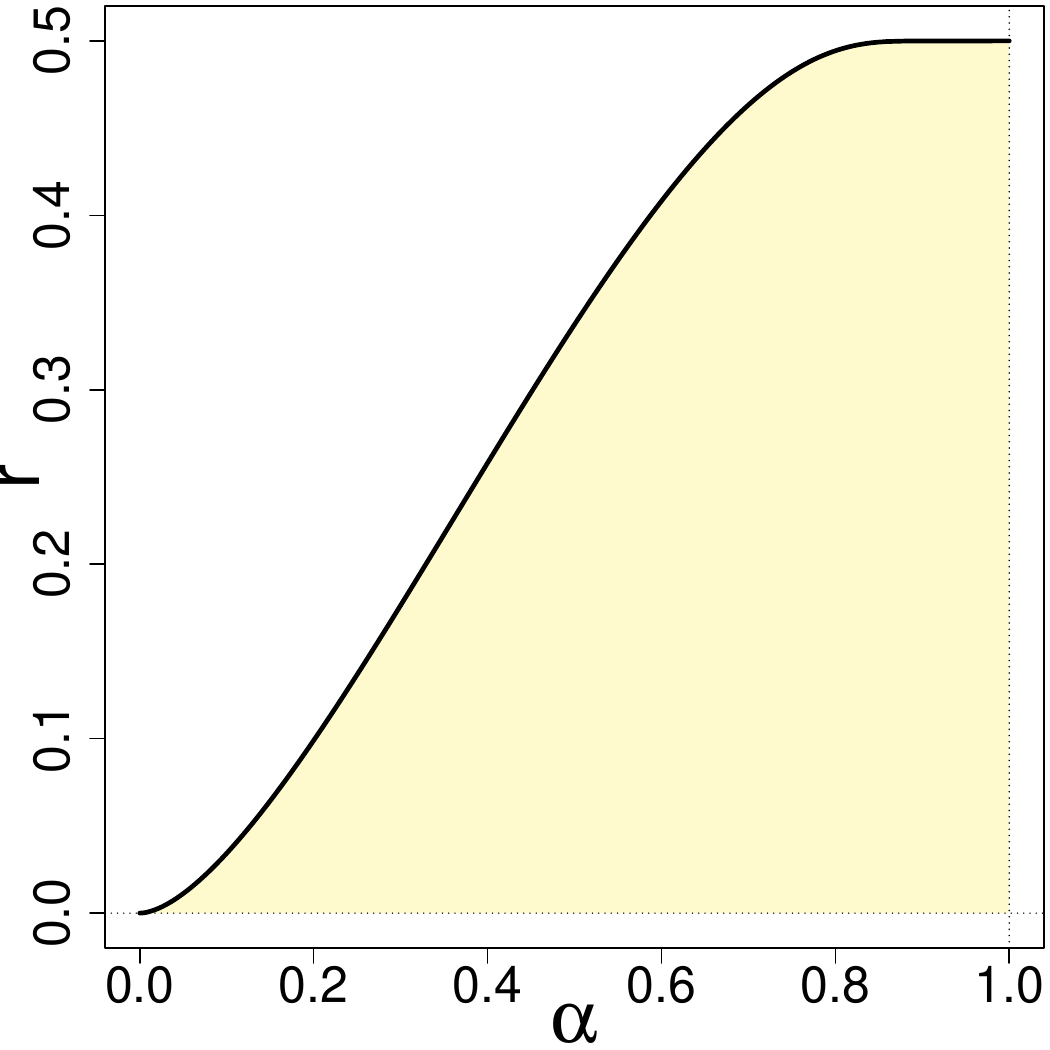}}
	\caption{\footnotesize The phase boundary of unregularized ES for i.i.d. normal underlying returns. 
In the region below the phase boundary the optimization of ES is feasible and the estimation error is finite. 
Approaching the phase boundary from below, the estimation error diverges, and above the line optimization is no longer feasible.}
	\label{figCriticalLine}
\end{figure}

The fundamental cause of the divergence of estimation error in the original, non-regularized problem
is that ES as a risk measure is not bounded from below. In finite samples it can happen that one of
the assets, or a combination of assets, dominates the others (i.e., produces a larger return than
the others in the given sample), thereby leading to an apparent arbitrage: one can achieve an
arbitrarily large gain (an arbitrarily large negative ES) by going very long in the dominant asset
and correspondingly short in the others \cite{ciliberti2007On,kondor2010Instability}. It is evident that this
apparent arbitrage is a mere statistical fluctuation, but along a special curve in the $r - \alpha$
plane this divergence occurs with probability one \cite{ciliberti2007On}. As a result, the
estimation error will diverge along the phase boundary shown in Fig.~\ref{figCriticalLine}. Note that in 
high-dimensional statistics where regularization is routinely applied the loss is always bounded
both from above and below. The setting in the present paper is, therefore, very different from 
the customary setup, which explaines the unusual results.

The purpose of regularization is to penalize the large fluctuations of the weight
vector, thereby eliminating this phase transition. 

Since ES is a piecewise linear function of the weights, the
quadratic regularizer will overcome excessive fluctuations, no matter how small the coefficient
$\eta$ is. Deep inside the region of stability (shown by pale yellow in Fig.\ref{figCriticalLine}), a weak regularizer
(small $\eta$) will modify the behavior of various quantities very little. In contrast, close to the
phase boundary, and especially in the vicinity of the point $\alpha = 1$, $r=0.5$ , where the solution
has an essential singularity, the effect of even a small $\eta$ is very strong, and beyond the
yellow region, where originally there was no solution, the regularizer will dominate the scene. In
the region where the solution is stable even without the regularizer, $r=N/T$ is small, which means we have
an abundance of data. We call this region the data-dominated region. In the presence of the
regularizer we will find finite solutions also far beyond the phase boundary, but here the regularizer 
dominates the data, so we can call this domain the bias-dominated region.

\section{Results}
The solution of the stationarity conditions can be obtained with the help of a computer.
In the following, we present the numerical solutions for the relative estimation error. The results will be 
displayed by constructing the contour map of this quantity, which will allow us to make a direct 
comparison between our present results and those in \cite{caccioli2015Portfolio}. 

\begin{figure}[H]
	\centerline{\includegraphics[width=7cm]{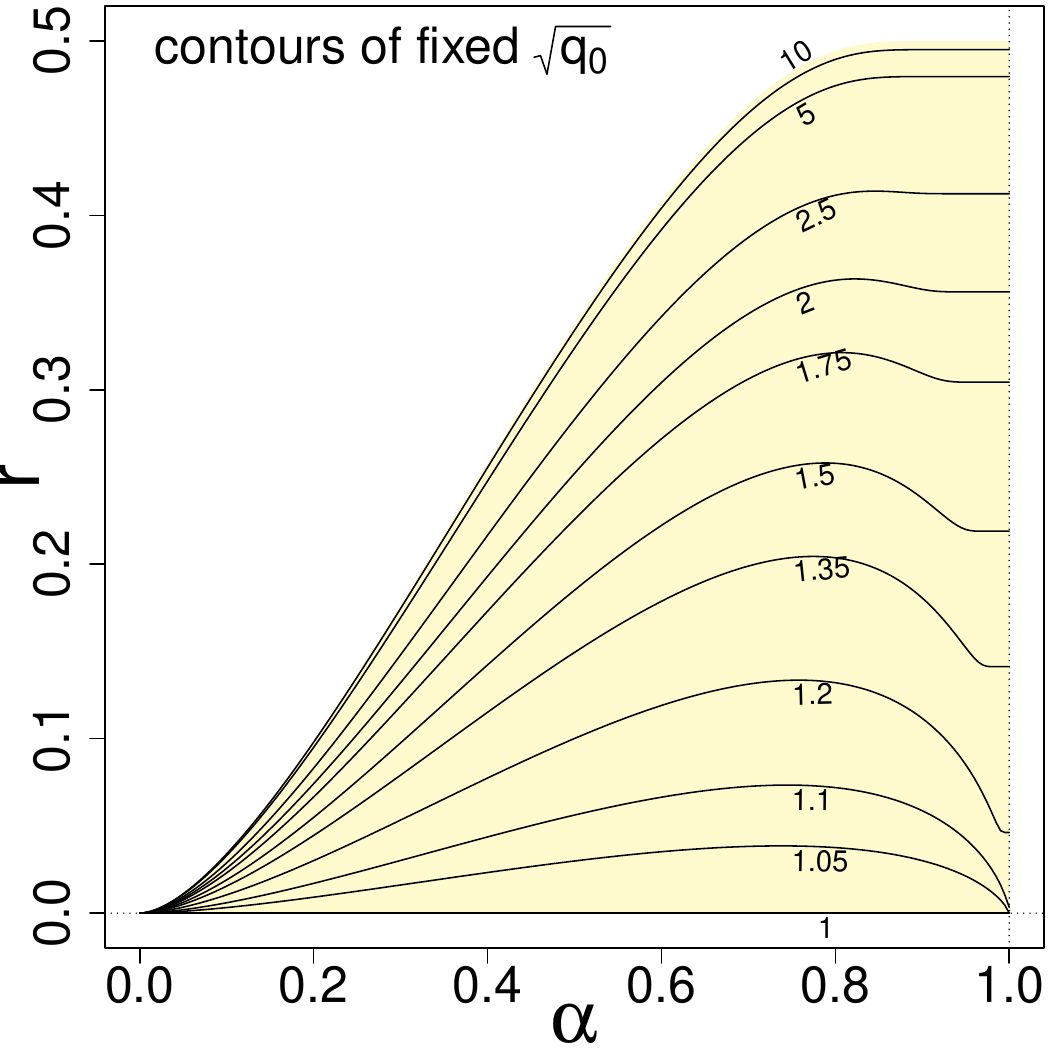}}
	\caption{\footnotesize Contour lines of fixed $\sqrt{q_0}$ in the absence of regularization. These 
curves are also the contour lines for the relative error for the out-of-sample estimate of ES.}
	\label{figEtaEvolr}
\end{figure}

In Fig.~\ref{figEtaEvolr} we recall the contour map of the relative estimation error of ES without regularization.

As can be seen, without regularization the constant $q_0$ curves are all inside the feasible
region. For larger and larger values of $q_0$ the corresponding curves run closer and closer to the
phase boundary, along which the estimation error diverges. Note that the phase boundary becomes
flat, with all its derivatives vanishing, at the upper right corner of the feasible region: there is an essential singularity at the
point $\alpha=1, r=0.5$.

The estimation error problem is very clearly illustrated in this figure: the curves corresponding to
an acceptably small relative error are the lowest ones among the $q_0$ contour lines, and the value
of $r=N/T$ corresponding to a confidence limit $\alpha$ in the vicinity of 1 (such as the regulatory
value $\alpha=0.975$) are extremely small on these low lying curves. These small values of $r$
require an unrealistically large sample size $T$ if $N$ is not small.  For example, at the regulatory
value of $\alpha=0.975$, to achieve an estimation error smaller than $5\%$, for a portfolio with
$N=100$ assets one would need a time series of more than $7200$ data points
\cite{caccioli2015Portfolio}.

Let us see how regularization reorganizes the set of constant $q_0$ curves. Figs.\:3 and 4. display
these curves for two different values of the coefficient $\eta$ of the regularizer. (Notice the
logarithmic scale on the vertical axes in these figures.)

\begin{minipage}{72mm}
\begin{figure}[H]
	\centerline{\includegraphics[width=70mm]{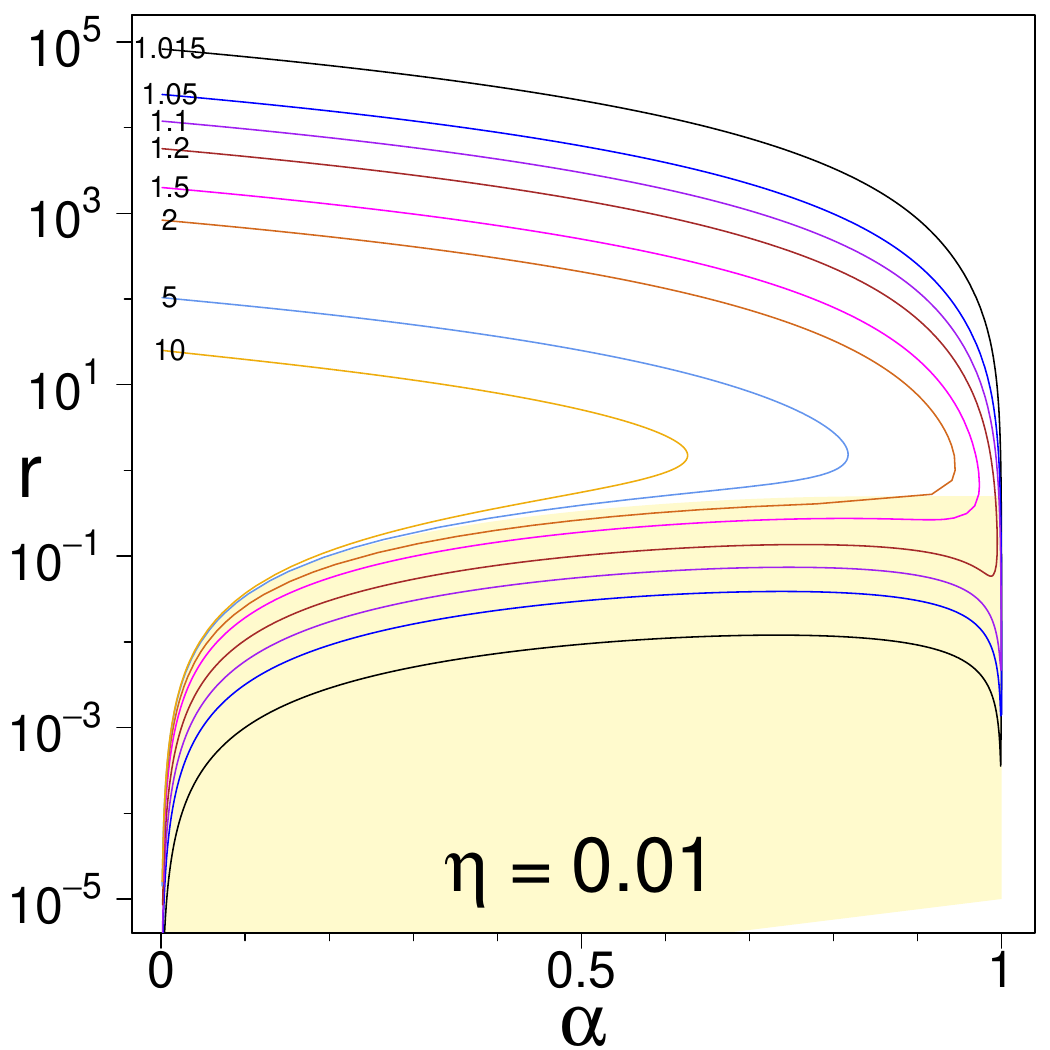}}
	\caption{\footnotesize Contour plot for fixed values of $\sqrt{q_0}$ on the $\alpha$ -- $r$ plane at $\eta=0.01$.}
	\label{figFixedq0_eta0.01}
\end{figure}
\end{minipage}
\hspace*{5mm}
\begin{minipage}{72mm}
\begin{figure}[H]
	\centerline{\includegraphics[width=70mm]{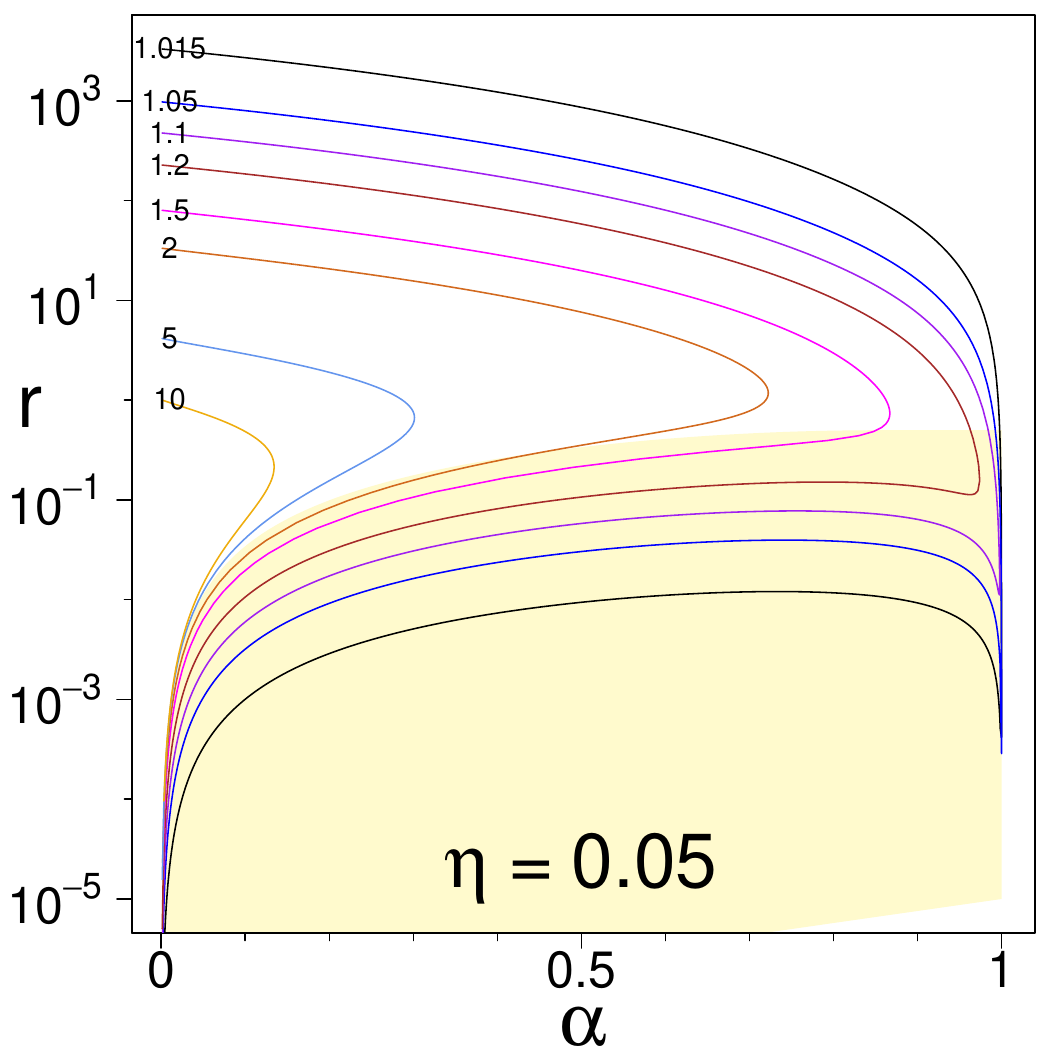}}
	\caption{\footnotesize Contour plot for fixed values of $\sqrt{q_0}$ on the $\alpha$ -- $r$ plane at $\eta=0.05$.}
	\label{figFixedq0_eta0.05}
\end{figure}
\end{minipage}

\bigskip

The curves of constant $q_0$ have two branches now. For a given $q_0$ the lower branch lies mostly
or partly in the previously feasible region, the upper branch lies outside, above it. Along the
lower branch the value of the ratio $r$ is small, which means we have very large samples with respect
to the dimension: this is the data-dominated regime. We can also see that when the data dominate,
the dependence on the regularizer is weak: the set of curves inside the yellow region is quite
similar in the two figures, even though the regularizer has been increased 5-fold
from Fig~\ref{figFixedq0_eta0.01} to Fig.~\ref{figFixedq0_eta0.05}. Following the curve corresponding to a given value of $q_0$, say the black
one, we see that at the beginning it is increasing with $\alpha$, in the vicinity of $\alpha=1$
it starts to decline, then it sharply turns around and shoots up steeply, leaving the feasible
region and increasing with decreasing $\alpha$. Along this upper branch the ratio $r$ is not small
any more. We do not have large samples here, in fact, the situation is just the opposite: the
dimension $N$ becomes larger than the size $T$ of the samples. Clearly, in this regime the
regularizer dominates and the data play only a minor role: this is the bias-dominated regime. 
It is interesting to note the sudden turn
over of the constant $q_0$ curves in the vicinity of $\alpha=1$. Such a sharp feature would be
extremely hard to discover if we wanted to solve the original optimization problem by numerical
simulations: the simulation would jump over to the upper branch before we could observe the sharp
dip and the identification of the boundary between the data-dominated and the bias-dominated 
regimes would be hard. This is even more so for real life data which are inevitably noisy.

An important point in regularization is the correct choice of the parameter $\eta$. When data come
from real observations, and the size of the sample (or the number of samples) is limited by time
and/or cost considerations, the standard procedure is cross validation \cite{hastie2008Elements},
i.e., using a part of the data to infer the value of $\eta$ and checking the correctness of this
choice on the other part. In the present analytical approach we have the luxury of infinitely many 
samples to average over, so we can obtain the value of the coefficient of regularization by demanding a given relative
error (that is a given $q_0$) for a given confidence limit $\alpha$ and given aspect ratio $r=N/T$.
\begin{figure}
	\centerline{\includegraphics[width=7cm]{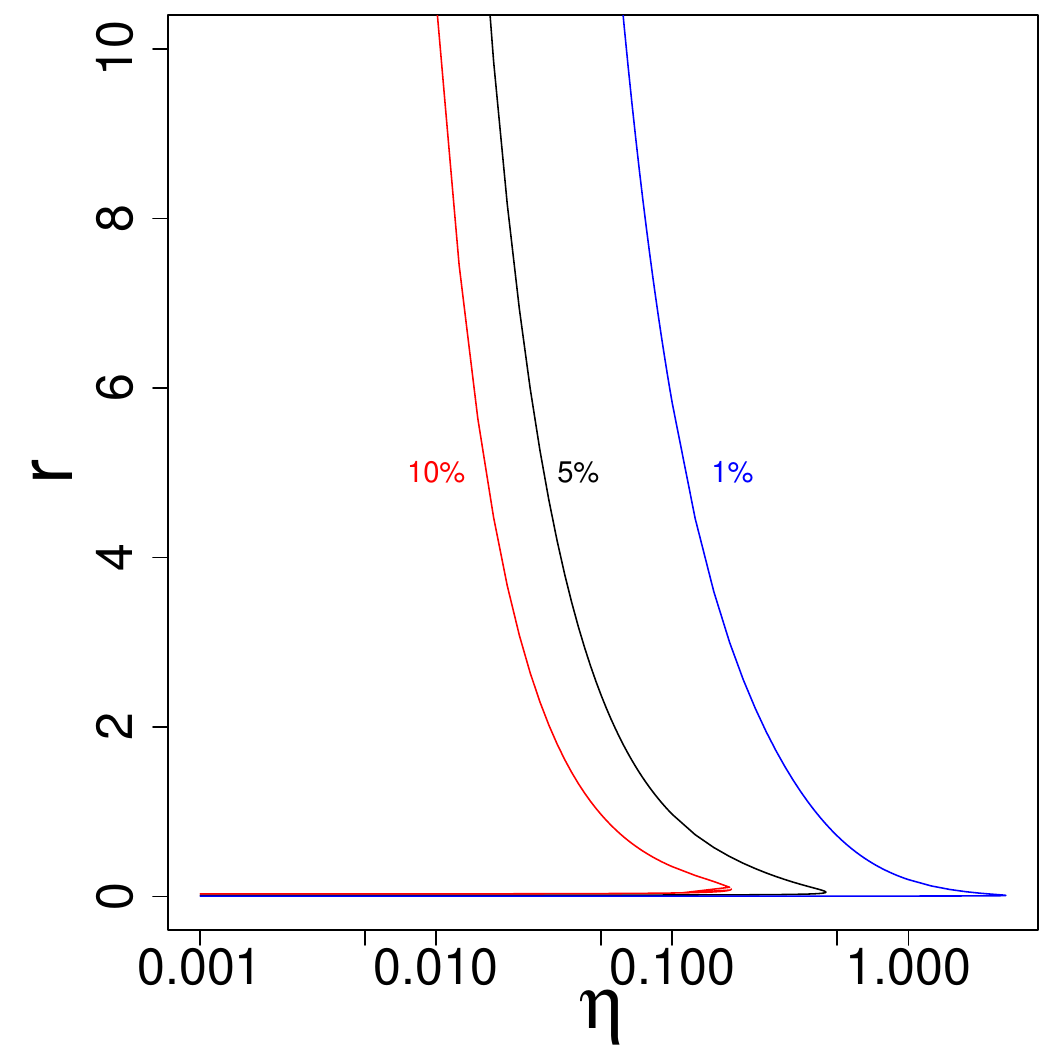}}
	\caption{\footnotesize The overall behavior of the contour lines of fixed estimation error (fixed $q_0$) on the $r=N/T$ - $\eta$ plane, for a given value of the confidence limit $\alpha=0.975$ and for three different values (1\%, 5\% and 10\%) of the relative estimation error. The data-dominated and bias-dominated regions correspond to the two branches of these curves: in the range of small $r$'s the curves depend on the strength of the regularizer very weakly, while for $r$'s in the vicinity of the phase boundary, and even more for large $r$'s high in the originally unfeasible region, the fixed estimation error curves display a strong dependence on $\eta$.}
	\label{figisoDelta_eta03}
\end{figure}
Fig.~\ref{figisoDelta_eta03} displays the plot of the given estimation error curves on the $r$ -\:$\eta$ plane for the specific 
value of $\alpha=0.975$ demanded by the new market risk
regulation \cite{basle2016Minimum}, and relative errors of 1\%, 5\% and 10\%, respectively. It shows
the change-over between the data-dominated resp.\:bias-dominated regimes very clearly. For a given
value of $r$ the corresponding value of $\eta$ can be read off from the curves. If $r$ is small
(i.e. the sample is large with respect to the dimension) the curves with the prescribed values of
relative error are almost horizontal. This means that when we have sufficient data the value of the
regularizer is more or less immaterial: within reasonable limits we can choose any coefficient for the
regularizer, it will not change the precision of the estimate, because in this situation the data
will determine the optimum. Conversely, when the data are insufficient ($r$ is not small, or it is
even beyond the feasible region), the value of $\eta$ necessary to enforce a given relative error strongly 
depends on $r$. In this region, however, we need a smaller and smaller $\eta$ to find the
same relative error, because here the data almost do not matter and even a small regularizer will
establish the optimum. The transition between these two regimes takes place around the points where
the curves turn back. This happens still inside the feasible region, and the width of this range is rather small: 
from the $r$ value corresponding to $\eta=0$ to the one where the curves turn around
the increase of $r$ always remains within a factor of about 4.

\begin{figure}
	\centerline{\includegraphics[width=7cm]{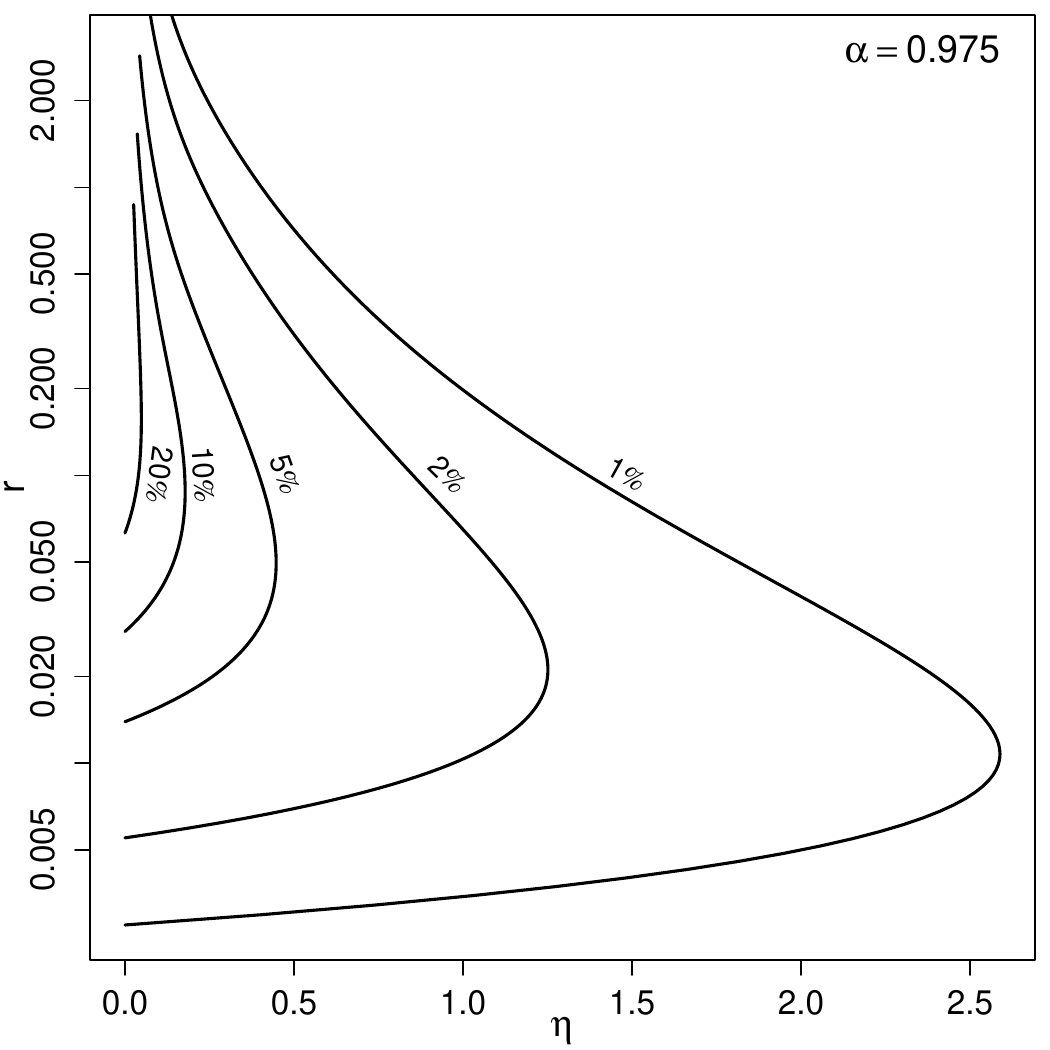} \hspace*{6mm} \includegraphics[width=7cm]{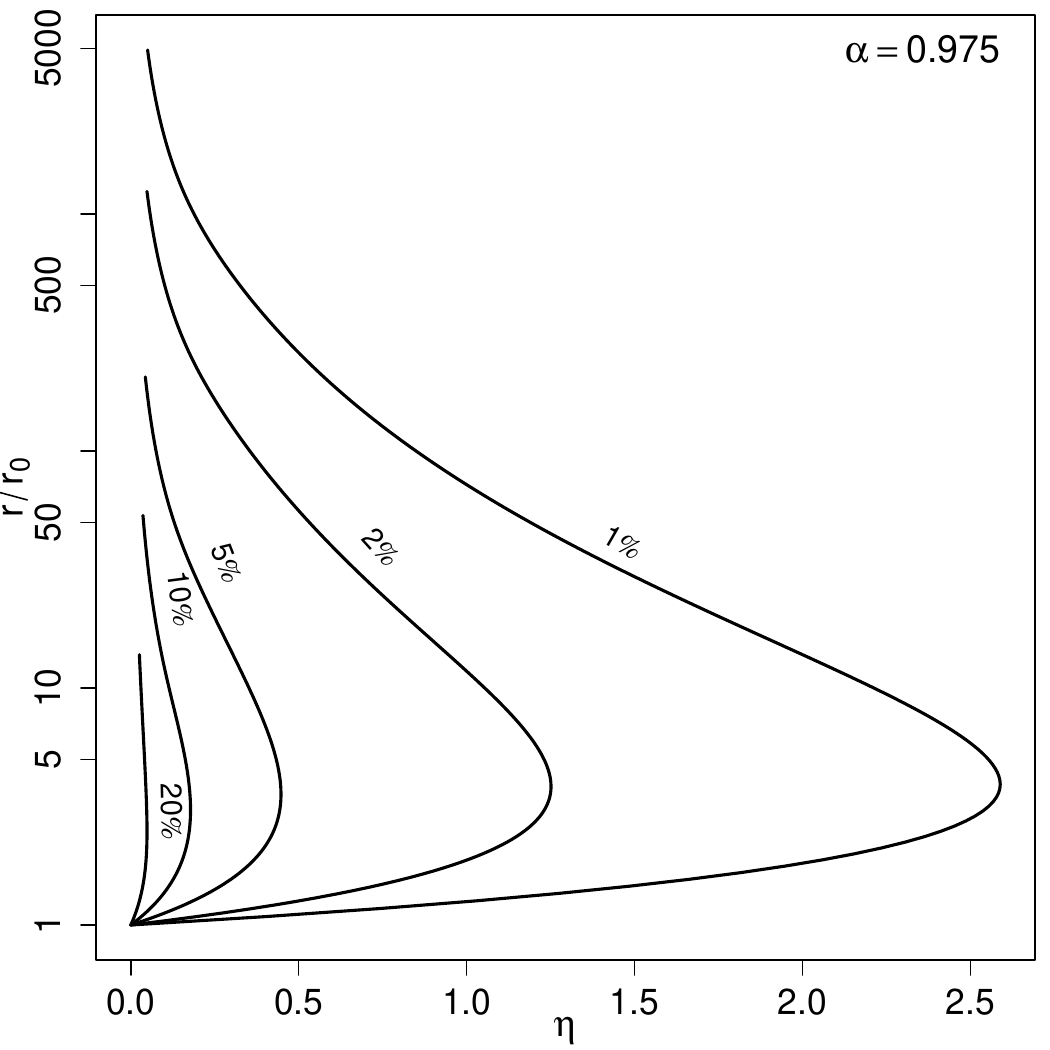}}
	\caption{\footnotesize {\bf a:} The previous figure in higher resolution (left). {\bf b:} The same as the left figure, but the $r(\eta)$ curves normalized by their initial values $r_0$ corresponding to $\eta=0$ (right). It can be seen that the gain in $r$ is about a factor 4.}
	\label{Fig_r_eta}
\end{figure}

Let us take a closer look at that part of the previous figure where the curves turn around and $r$ starts to increase. Fig.~\ref{Fig_r_eta}a shows this region in higher resolution. For a given, small, value of the estimation error (such as 1\% or 2\%), $r$ grows by a factor of about 4 by the time we reach the elbow of the curves (at rather large $\eta$ values). This means that for a given sample size $T$ the regularization allows us to consider a four times larger portfolio without increasing the estimation error. Conversely, for a given value of $N$ the regularizer allows the use of four times shorter time series without compromising the quality of the estimate. Of course, the growth of $r$ could be followed beyond the elbow, up to higher values along the constant $q_0$ curves, but it must be clear that these sections of the curves correspond to a situation where the estimate is mostly or entirely determined by the regularizer. This is also shown by the fact that the curves of given estimation error strongly lean backwards to the vertical axis: where the dimension is high and the data few even a weak regularizer can stabilize the estimate, but it will then have nothing to do with the information coming from the time series.

A gain of a factor 4 in the allowed region in $r$ could be regarded as very satisfactory, were it not for the fact that the initial ($\eta=0$) value of $r$ along the small estimation error curves is so small that it remains small even after a multiplication by 4.  

If we inspect another curve, corresponding to a larger estimation error (say, 5\%), we can see that it turns back for a much smaller $\eta$, but the relative increase of $r$ up to the elbow is still about a factor 4. We can also see that beyond this point the curves very quickly reach the region where the regularizer dominates.

Figure.~\ref{Fig_r_eta}b displays the same curves as in Fig.~\ref{Fig_r_eta}a, but this time they are normalized by their vales at $\eta=0$, so that they show the gain in $r$ due to the regularizer.

\section{Conclusion}

We have considered the problem of optimizing Expected Shortfall in the presence of an $\ell_2$
regularizer. The regularizer takes care of the large sample fluctuations and eliminates the phase
transition that would be present in the problem without regularization. Deep inside the feasible
region, where we have a large amount of data relative to the dimension, the size of the sample needed
for a given level of relative estimation error is basically constant, largely independent of the
regularizer. In the opposite case, for sample sizes comparable to or small relative to the
dimension, the regularizer dominates the optimization and suppresses the data. The transition
between the the data-dominated regime and the regularizer-dominated one is rather narrow. 
It is in this transition region where we can
meaningfully speak about a trade-off between fluctuation and bias, otherwise one or the other
dominates the estimation. The identification of this transitional zone is easy within the
present scheme, where we could carry out the optimization analytically: the transitional zone is the
small region where the curves in Fig.~\ref{figisoDelta_eta03} sharply turn back, but still remain inside the originally feasible region.
In real life, where the size of the
samples can rarely be changed at will and where all kinds of external noise (other than that
coming from the sample fluctuations) may be present, the distinction between the region where the
data dominates and where the bias reigns may be much less clear, and one may not be sure where the
transition takes place between them. Below this transition there is not much point in using
regularization, because the data themselves are sufficient to provide a stable and reliable
estimate. Above the transition zone it is almost meaningless to talk about the observed data, because
they are crowded out by the bias. The identification of the relatively narrow transition zone 
between these two extremes and the gain of a factor 4 below the transition are the main results of this paper. 

It is important to realize, however, that the cause of this narrow transition region is the same as that of the strong fluctuations, namely the unbounded loss function. Expected Shortfall is not the only risk measure to have this deficiency: all the downside risk measures have it, including Value at Risk. The preference for downside risk measures is explained by the fact that investors (and regulators) are not afraid of big gains, only of big losses. Perhaps they should be. Refusing to acknowledge the risk in improbably large gains is a Ponzi scheme mentality. Downside risk measures embody this mentality.  As a part of regulation, however, they acquire an air of undeserved respectability, at which point the associated technical issues become components of systemic risk.

\section{Acknowledgements}

We are obliged to R.\:Kondor, M.\:Marsili and S.\:Still for valuable discussions. FC acknowledges
support of the Economic and Social Research Council (ESRC) in funding the Systemic Risk Centre
(ES/K002309/1). IK is grateful for the hospitality extended to him at the Computer Science
Department of the University College of London where this paper was written up.

\bibliography{ContourLines}

\end{document}